\documentclass[graybox]{svmult}

\usepackage{helvet}         
\usepackage{courier}        
\usepackage{type1cm}        
\usepackage{makeidx}         
\usepackage{graphicx}        
\usepackage{multicol}        
\usepackage[bottom]{footmisc}

\usepackage{amsmath}         
\usepackage{amssymb}         
\usepackage{cite}            
\usepackage[colorlinks=true,urlcolor=blue,linkcolor=blue]{hyperref}  

\begin{document}

\title*{Nonholomorphic superpotentials in heterotic Landau-Ginzburg models}
\author{Richard S. Garavuso}
\institute{Richard S. Garavuso \at Kingsborough Community College, The City University of New York, 2001 Oriental Boulevard, Brooklyn, NY 11235-2398, USA, \email{richard.garavuso@kbcc.cuny.edu}}
\maketitle

\abstract{
The aim of this talk is to derive two constraints imposed by supersymmetry for a class of heterotic Landau-Ginzburg models with nonholomorphic superpotentials.
One of these constraints relates the nonholomorphic parameters of the superpotential to the
Hermitian curvature.
Various special cases of this constraint have been used to establish properties of Mathai-Quillen form analogues which arise in the corresponding heterotic Landau-Ginzburg models.
The other constraint was not anticipated from studies of Mathai-Quillen form analogues.
         }

\vskip 2ex
Invited talk presented at the 14th International Workshop ``Lie Theory and Its Applications in Physics'' (LT-14), 21-25 June 2021, Sofia, Bulgaria (online).
\numberwithin{equation}{section}

\section{\label{Intro}Introduction}

A Landau-Ginzburg model is a nonlinear sigma model with a superpotential.
For a \emph{heterotic} Landau-Ginzburg model
\cite{Witten:Phases, DistlerKachru:0-2-Landau-Ginzburg, AdamsBasuSethi:0-2-Duality, MelnikovSethi:Half-twisted, GuffinSharpe:A-twistedheterotic, MelnikovSethiSharpe:Recent-Developments, GaravusoSharpe:Analogues, Garavuso:Curvaure-constraints}, 
the nonlinear sigma model possesses only
$ (0, 2) $
supersymmetry and the superpotential is a Grassmann-odd function of the superfields which may or may not be holomorphic.
It was claimed in
\cite{GaravusoSharpe:Analogues}
that, for various heterotic Landau-Ginzburg models with nonholomorphic superpotentials,
supersymmetry imposes a constraint which relates the nonholomorphic parameters of the superpotential to the Hermitian curvature.
Details supporting that claim were worked out in
\cite{Garavuso:Curvaure-constraints}.
The analysis revealed an additional constraint imposed by supersymmetry which was not anticipated in
\cite{GaravusoSharpe:Analogues}.  
This talk will summarize the analysis found in
\cite{Garavuso:Curvaure-constraints}.

This talk is organized as follows:
In section 
\ref{section:Action}, 
we will write down the action for the class of heterotic Landau-Ginzburg models that we are considering.
In section
\ref{section:Supersymmetry-holomorphic}, 
for the case of a holomorphic superpotential, we demonstrate that the action is invariant on-shell under supersymmetry transformations up to a total derivative.
Finally, in section
\ref{section:Supersymmetry-nonholomorphic},
we will extend the analysis to the case in which the superpotential is not holomorphic.
In this case, we obtain two constraints imposed by supersymmetry.
\section{\label{section:Action}Action}

Heterotic Landau-Ginzburg models have field content consisting of
$ (0, 2) $ 
bosonic chiral superfields
$ \varPhi^i = (\phi^i, \psi^i_+ )$
and
$ (0, 2) $ 
fermionic chiral superfields 
$ 
  \varLambda^a 
    = \left( \lambda^a_-, 
             H^a, 
             E^a
      \right) 
$,
along with their conjugate antichiral superfields
$ 
  \varPhi^{\overline{\imath}} 
    = \left(
             \phi^{\overline{\imath}}, 
             \psi^{\overline{\imath}}_+ 
      \right)
$
and
$ 
  \varLambda^{\overline{a}} 
    = \left( 
             \lambda^{\overline{a}}_-, 
             \overline{H}^{\overline{a}}, 
             \overline{E}^{\overline{a}}
      \right) 
$.
The 
$ \phi^i $ 
are local complex coordinates on a K{\"a}hler manifold 
$ X $. 
The 
$ E^a $
are local smooth sections of a Hermitian vector bundle 
$ \mathcal{E} $
over 
$ X $,
i.e. 
$ E^a \in \varGamma(X, \mathcal{E}) $. 
The
$ H^a $
are nonpropagating auxiliary fields.
The fermions couple to bundles as follows:
\begin{align*}
\psi^i_+ 
  \in \varGamma
      \left(
             K^{1/2}_{\varSigma}
             \otimes
             \varPhi^* \!
             \left(
                    T^{1,0} 
                    X
             \right)      
      \right),
\qquad
&
\lambda^a_- 
  \in \varGamma
      \left(
             \overline{K}^{1/2}_{\varSigma}
             \otimes
             \left(
                    \varPhi^* 
                    \overline{\mathcal{E}}
             \right)^{\vee}
      \right),
\\
\psi^{\overline{\imath}}_+
  \in \varGamma
      \left(
             K^{1/2}_{\varSigma}
             \otimes
             \left(
                    \varPhi^* \!
                    \left(
                           T^{1,0}
                           X
                    \right)       
             \right)^{\vee}
      \right),
\qquad
&
\lambda^{\overline{a}}_-
  \in \varGamma
      \left(
             \overline{K}^{1/2}_{\varSigma}
             \otimes
             \varPhi^* 
             \overline{\mathcal{E}}
      \right),                    
\end{align*}
where
$ \varPhi: \varSigma \rightarrow X $
and
$ K_{\varSigma} $
is the canonical bundle on the worldsheet
$ \varSigma $.

In 
\cite{GuffinSharpe:A-twistedheterotic},
heterotic Landau-Ginzburg models with superpotential of the form
\begin{equation}
  W = \varLambda^a \,
      F_a \, ,
\end{equation}      
where 
$ 
  F_a 
  \in
  \varGamma
  \left(
         X, 
         \mathcal{E}^{\vee}  
  \right)
$,
were considered.
In this talk, we will study supersymmetry in these heterotic Landau-Ginzburg models with 
$ E^a = 0 $.
Assume that
$ X $ 
has metric
$ g $, 
antisymmetric tensor
$ B $,
and local real coordinates
$ \phi^{\mu} $.
Furthermore, assume that
$ \mathcal{E} $
has Hermitian fiber metric
$ h $.
Then the action of a Landau-Ginzburg model on 
$ X $
with gauge bundle
$ \mathcal{E} $
and
$ E^a = 0 $
can be written
\cite{Garavuso:Curvaure-constraints}
\begin{align}
\label{action}
S &= 2t \int_{\varSigma} d^2 z 
     \left[ 
            \frac{1}{2} 
            \left( 
                   g_{\mu \nu} + i B_{\mu \nu}
            \right)
            \partial_z 
            \phi^{\mu} 
            \partial_{\overline{z}} 
            \phi^{\nu}
    \right.        
\nonumber
\\
  &\phantom{= 2t \int_{\varSigma} d^2 z \left[ \right.}            
          + i 
            g_{\overline{\imath} i}
            \psi_+^{\overline{\imath}}
            \overline{D}_{\overline{z}} \psi_+^i
          + i 
            h_{a \overline{a}} 
            \lambda_-^a
            D_z \lambda_-^{\overline{a}}
          + F_{i \overline{\imath} a \overline{a}} \,
            \psi_+^i
            \psi_+^{\overline{\imath}}
            \lambda_-^a
            \lambda_-^{\overline{a}}
\nonumber
\\
  &\phantom{= 2t \int_{\varSigma} d^2 z \left[ \right.}
     \biggl. \,
   + \, h^{a \overline{a}} 
     F_{a} 
     \overline{F}_{\overline{a}}   
   + \psi_+^i 
     \lambda_-^{a} 
     D_i F_a 
   + \psi_+^{\overline{\imath}}
     \lambda_-^{\overline{a}} 
     \overline{D}_{\overline{\imath}} \overline{F}_{\overline{a}}
     \biggr].     
\end{align}
Here,
$ t $ 
is a coupling constant,
$ d^2 z = -i \, dz \wedge d{\overline{z}} $,
and
\begin{align*}
\overline{D}_{\overline{z}} \,
\psi^i_+
   &= \overline{\partial}_{\overline{z}} \,
      \psi_+^i
    + \overline{\partial}_{\overline{z}} \,
      \phi^j \,
      \varGamma^i_{jk}
      \psi^k_+ \, ,
&D_z \lambda^{\overline{a}}_-
  &= \partial_z 
     \lambda_-^{\overline{a}}
   + \partial_z \phi^{\overline{\imath}}
     A^{\overline{a}}_{\overline{\imath} \overline{b}} \,
     \lambda^{\overline{b}}_- \, ,
\\
D_i F_a
  &= \partial_i F_a
   - A^{b}_{i a}
     F_b \, ,
& \overline{D}_{\overline{\imath}}
  \overline{F}_{\overline{a}}
  &= \overline{\partial}_{\overline{\imath}} \,
     \overline{F}_{\overline{a}}
   - A^{\overline{b}}_{\overline{\imath} \, \overline{a}} \,
     \overline{F}_{\overline{b}} \, ,
\\
A^b_{i a} 
  &= h^{b \overline{b}} \,
     h_{\overline{b} a, i} \, ,
& A^{\overline{b}}_{\overline{\imath} \, \overline{a}} 
  &= h^{\overline{b} b} \, 
     h_{b \overline{a}, \overline{\imath}} \, ,     
\\
\varGamma^i_{jk}
  &= g^{i \overline{\imath}} \,
     g_{\overline{\imath} k, j} \, ,         
& F_{i \overline{\imath} a \overline{a}}  
   &= h_{a \overline{b}} \,
      A^{\overline{b}}_{\overline{\imath} \, \overline{a}, i} \, .
\end{align*}

The action 
\eqref{action}
is invariant on-shell under the supersymmetry transformations
\begin{equation}
\begin{aligned}
\delta \phi^i
  &= i 
     \alpha_-
     \psi^i_+ \, ,
\\
\delta \phi^{\overline{\imath}}
  &= i 
     \tilde{\alpha}_-
     \psi^{\overline{\imath}}_+ \, ,          
\\
\delta \psi^i_+ 
  &= 
   - \tilde{\alpha}_-
     \overline{\partial}_{\overline{z}} \phi^i \, ,
\\
\delta \psi^{\overline{\imath}}_+
  &=
   - \alpha_-
     \partial_z \phi^{\overline{\imath}} \, ,
\\
\delta \lambda^{a}_-
  &= 
   - i
     \alpha_-
     \psi^j_+ \,
     A^a_{j b} \,
     \lambda^{b}_- 
   + i
     \alpha_-
     h^{a \overline{a}} \,
     \overline{F}_{\overline{a}} \, ,
\\
\delta \lambda^{\overline{a}}_-
  &= 
    - i
     \tilde{\alpha}_-
     \psi^{\overline{\jmath}}_+ \,
     A^{\overline{a}}_{\overline{\jmath} \, \overline{b}} \,
     \lambda^{\overline{b}}_- 
   + i
     \tilde{\alpha}_-
     h^{\overline{a} a} \,
     F_a                 
\end{aligned}
\label{SUSY}
\end{equation}
up to a total derivative.
\section{\label{section:Supersymmetry-holomorphic}Supersymmetry invariance for holomorphic superpotential}

In this section, we will show that, when the superpotential is holomorphic, the action
\eqref{action}
is invariant on shell under the supersymmetry transformations
\eqref{SUSY}
up to a total derivative.
For this purpose, it is sufficient to set
$ \tilde{\alpha}_- = 0 $,\footnote{
The calculations for the case in which
$ \alpha_- = 0 $ 
and 
$ \tilde{\alpha}_- \neq 0 $
are analogous to those we will perform explicitly for the case in which
$ \alpha_- \neq 0 $ 
and 
$ \tilde{\alpha}_- = 0 $.
The general case, i.e.
$ \alpha_- $
and
$ \tilde{\alpha}_- $
both nonzero, is obtained by combining the above two cases.
                                }
yielding
\begin{equation}
\begin{aligned}
\delta \phi^i
  &= i 
     \alpha_-
     \psi^i_+ \, ,
\qquad      
&\delta \phi^{\overline{\imath}}
  &= 0 \, ,      
\\
\delta \psi^i_+ 
  &= 0 \, ,
\qquad  
&\delta \psi^{\overline{\imath}}_+
  &=
   - \alpha_-
     \partial_z \phi^{\overline{\imath}} \, ,
\\
\delta \lambda^{a}_-
  &= 
   - i
     \alpha_-
     \psi^j_+ \,
     A^a_{j b} \,
     \lambda^{b}_- 
   + i
     \alpha_-
     h^{a \overline{a}} \,
     \overline{F}_{\overline{a}} \, ,
\qquad
&\delta \lambda^{\overline{a}}_-
  &= 0 \, .         
\end{aligned}
\end{equation}
With this simplification, using the 
$ \lambda^a_- $ 
equation of motion,\footnote{This is valid because we have integrated out the auxiliary fields $ H^a $.}
the action
\eqref{action} 
can be written
\cite{Garavuso:Curvaure-constraints}
\begin{align}
\label{action:Q-exact+non-exact}
S &= it \int_{\varSigma} d^2 z \,
     \left\{
             Q, V
     \right\}
   + t \int_{\varSigma}
      \varPhi^*(K)
\nonumber
\\      
  &\phantom{=}
   + 2 t 
     \int_{\varSigma} d^2 z
     \left(
            \psi_+^{\overline{\imath}}
            \lambda_-^{\overline{a}} 
            \overline{D}_{\overline{\imath}} \overline{F}_{\overline{a}}
          - \psi_+^i 
            \lambda_-^{a} 
            D_a F_a
     \right),               
\end{align}
where
$ Q $
is the BRST operator, 
\begin{equation}
V = 2 \left(
               g_{\overline{\imath} i}
               \psi^{\overline{\imath}}_+ 
               \overline{\partial}_{\overline{z}} \phi^i
             + i 
               \lambda^a_- 
               F_a         
      \right),      
\end{equation}
and 
\begin{equation}
\int_{\varSigma} \varPhi^*(K) 
   = \int_{\varSigma} d^2 z 
     \left(
            g_{i \overline{\imath}}
          + i 
            B_{i \overline{\imath}}  
     \right)
     \left(
            \partial_z \phi^i \,
            \overline{\partial}_{\overline{z}} \phi^{\overline{\imath}}
          - \overline{\partial}_{\overline{z}} \phi^i
            \partial_z \phi^{\overline{\imath}} 
     \right)          
\end{equation}
is the integral over the worldsheet 
$ \varSigma $
of the
pullback to
$ \varSigma $
of the complexified K{\"a}hler form
$ 
  K = 
    - i 
      \left(
             g_{i \overline{\imath}}
           + i
             B_{i \overline{\imath}} 
      \right)       
             d \phi^i
             \wedge
             d \phi^{\overline{\imath}}  
$.

Since 
$ \delta f = -i \alpha_- \{ Q, f \} $, 
where 
$ f $ 
is any field, the 
$ Q $-exact part of
\eqref{action:Q-exact+non-exact}
is
$ \delta $-exact 
and hence
$ \delta $-closed.
We will now establish that the remaining terms of
\eqref{action:Q-exact+non-exact}
are
$ \delta $-closed 
on shell up to a total derivative.
For the non-exact term of
\eqref{action:Q-exact+non-exact}
involving
$ \varPhi^*(K) $,
note that
\begin{equation*}
\int_{\varSigma}
\varPhi^*(K)
  = \int_{\varPhi(\varSigma)} K
  = \int_{\varPhi(\varSigma)}
    \left[
         - i 
           \left(
                  g_{i \overline{\imath}}
                + i
                  B_{i \overline{\imath}} 
           \right)
    \right]       
    d \phi^i
    \wedge
    d \phi^{\overline{\imath}}
\end{equation*}
and
$ K $
satisfies
\begin{equation*}
\partial K
 =
 - i \, 
   \partial_k
   \left(
          g_{i \overline{\imath}}
        + i
          B_{i \overline{\imath}} 
   \right)
    d \phi^k
    \wedge
    d \phi^i
    \wedge
    d \phi^{\overline{\imath}}
 = 0 \, .    
\end{equation*}
Thus,
\begin{equation}
\delta
\left[ 
       \varPhi^*(K)
\right]       
  = \left[
           \varPhi^*(K)
    \right]_k
    \delta
    \phi^k
  = 0 \, .         
\end{equation}
It remains to consider the non-exact expression of
\eqref{action:Q-exact+non-exact}
involving
$$
  \psi_+^{\overline{\imath}}
  \lambda_-^{\overline{a}} 
  \overline{D}_{\overline{\imath}} \overline{F}_{\overline{a}}
- \psi_+^i 
  \lambda_-^{a} 
  D_i F_a \, .
$$
First, we compute 
\begin{align}
\label{delta-non-exact-2a}
\delta
\left(
       \psi^{\overline{\imath}}_+      
       \lambda^{\overline{a}}_-       
       \overline{D}_{\overline{\imath}}
       \overline{F}_{\overline{a}}
\right)
  &= \left(
            \delta \psi_+^{\overline{\imath}} 
     \right)       
     \lambda^{\overline{a}}_- \,
     \overline{D}_{\overline{\imath}} \,
     \overline{F}_{\overline{a}}
   + \psi^{\overline{\imath}}_+
     \left(
            \delta \lambda^{\overline{a}}_-
     \right)       
     \overline{D}_{\overline{\imath}} \,
     \overline{F}_{\overline{a}}
   + \psi^{\overline{\imath}}_+
     \lambda^{\overline{a}}_i
     \left[
            \delta \!
            \left( \,
                   \overline{D}_{\overline{\imath}} \,
                   \overline{F}_{\overline{a}}
            \right) 
     \right]       
\allowdisplaybreaks
\nonumber
\\[1ex]
  &= \left(
          - \alpha_-
            \partial_z \phi^{\overline{\imath}}
     \right)       
     \lambda_-^{\overline{a}} \,
     \overline{D}_{\overline{\imath}} \,
     \overline{F}_{\overline{a}}
   + \psi^{\overline{\imath}}_+
     \lambda^{\overline{a}}_-
     \left[
            \delta
            \left(
                   \overline{\partial}_{\overline{\imath}}
                   \overline{F}_{\overline{a}}
                 - A^{\overline{b}}_{\overline{\imath} \, \overline{a}} \,
                   \overline{F}_{\overline{b}}   
            \right)
     \right]
\nonumber
\\[1ex]
  &= \left(
          - \alpha_-
            \partial_z \phi^{\overline{\imath}}
     \right)       
     \lambda_-^{\overline{a}} \,
     \overline{D}_{\overline{\imath}} \,
     \overline{F}_{\overline{a}}     
   + \psi^{\overline{\imath}}_+
     \lambda_-^{\overline{a}}
     \left\{
            \overline{\partial}_{\overline{\imath}}
            \left[ \, 
                   \overline{F}_{\overline{a},k}
                   \left( 
                          \delta \phi^k
                   \right)       
            \right]
     \right.
\nonumber
\\     
  &\phantom{=}
     \left.                   
          - \left[ 
                   A^{\overline{b}}_{\overline{\imath} \, \overline{a}, k}
                   \left( 
                          \delta \phi^k
                   \right)                   
            \right]
            \overline{F}_{\overline{b}}
          - A^{\overline{b}}_{\overline{\imath} \, \overline{a}}
            \left[
                   \overline{F}_{\overline{b}, k}
                   \left(
                          \delta \phi^k
                   \right) 
            \right]  
     \right\}
\allowdisplaybreaks
\nonumber
\\[1ex]
  &= \left(
          - \alpha_-
            \partial_z \phi^{\overline{\imath}}
     \right)       
     \lambda^{\overline{a}}_- \,
     \overline{D}_{\overline{\imath}} \,
     \overline{F}_{\overline{a}}      
   - \psi_+^{\overline{\imath}}
     \lambda_-^{\overline{a}}
     A^{\overline{b}}_{\overline{\imath} \, \overline{a}, k}
     \left( 
            i
            \alpha_-
            \psi^k_+                   
     \right)                   
     \overline{F}_{\overline{b}} \, ,                         
\end{align}
where we have used
$ \overline{F}_{\overline{a},k}  = 0 $
in the last step.
Now, we compute
\begin{align}
\label{delta-non-exact-2b}
\delta
\left(
    - \psi^i_+ 
       \lambda^a_-
       D_i F_a
\right)
  &= 
   - \,
     \left(
            \delta \psi^i_+
     \right)       
     \lambda^a_- 
     D_i F_a
   - \psi^i_+
     \left(
            \delta \lambda^a_-
     \right)        
     D_i F_a
\nonumber
\\
  &\phantom{\, = \,}
   - \psi^i_+
     \lambda^a_-
     \left[
            \delta \!
            \left( 
                   D_i F_a
            \right)
     \right]       
\nonumber     
\\
  &=
   - \,
     \alpha_-
     \overline{F}_{\overline{a}}
     D_z \lambda^{\overline{a}}_-
   + \left(
            i
            \alpha_-
            h^{a \overline{b}} \,
            \overline{F}_{\overline{b}}
     \right)    
     F_{i \overline{\imath} a \overline{a}} \,
     \psi^i_+
     \psi^{\overline{\imath}}_+
     \lambda^{\overline{a}}_- \, ,                                                    
\end{align}
where we have used the 
$ \lambda^a_- $
equation of motion.
Note that the first term on the right-hand side of
\eqref{delta-non-exact-2b}
cancels the first term on the right-hand side of
\eqref{delta-non-exact-2a}
up to a total derivative: 
\begin{align}
\label{cancel-first-terms-rhs-nonexact-2a-2b}
- \,
    \alpha_-
    \overline{F}_{\overline{a}} \,
    D_z \lambda^{\overline{a}}_-
 &=
  - \,
    \alpha_-
    \overline{F}_{\overline{a}}
    \left(
           \partial_z \lambda^{\overline{a}}_-
         + \partial_z \phi^{\overline{\imath}} \,
           A^{\overline{a}}_{\overline{\imath} \, \overline{b}} \,
           \lambda^{\overline{b}}_-
    \right)
\nonumber
\\[1ex]
  &= \alpha_-
     \left( 
            \overline{F}_{\overline{a}, k} \,
            \partial_z \phi^k
          + \overline{F}_{\overline{a}, \overline{k}} \,
            \partial_z \phi^{\overline{k}} 
     \right)
     \lambda^{\overline{a}}_-
   - \alpha_-
     \partial_z \!
     \left(
            \overline{F}_{\overline{a}} \,
            \lambda^{\overline{a}}_-            
     \right)
\nonumber
\\
  &\phantom{=}     
   - \alpha_-
     \overline{F}_{\overline{a}} \,
     \partial_z \phi^{\overline{\imath}} \,
     A^{\overline{a}}_{\overline{\imath} \, \overline{b}} \,
     \lambda^{\overline{b}}_- 
\nonumber
\\[1ex]
 &= \left(
           \alpha_-
           \partial_z \phi^{\overline{\imath}}
    \right)       
    \lambda^{\overline{a}}_-
    \left(
           \overline{\partial}_{\overline{\imath}}
           \overline{F}_{\overline{a}}
         - A^{\overline{b}}_{\overline{\imath} \, \overline{a}} \,
           \overline{F}_{\overline{b}}   
    \right)
  - \alpha_-
    \partial_z \!
    \left(
           \overline{F}_{\overline{a}} \,
           \lambda^{\overline{a}}_-
     \right)   
\nonumber
\\[1ex]
 &= \left(
           \alpha_-
           \partial_z \phi^{\overline{\imath}}
    \right)       
    \lambda^{\overline{a}}_- \,
    \overline{D}_{\overline{\imath}} \,
    \overline{F}_{\overline{a}}
  - \alpha_-
    \partial_z \!
    \left(
           \overline{F}_{\overline{a}} \,
           \lambda^{\overline{a}}_-
    \right),                            
\end{align}
where we used
$ \overline{F}_{\overline{a}, k} = 0 $
in the third step.
Furthermore, the second term
on the right-hand side of
\eqref{delta-non-exact-2b}
cancels the second term
on the right-hand side of
\eqref{delta-non-exact-2a}:
\begin{align}
\left(
       i
       \alpha_-
       h^{a \overline{b}} \,
       \overline{F}_{\overline{b}}
\right)    
F_{i \overline{\imath} a \overline{a}} \,
\psi^i_+
\psi^{\overline{\imath}}_+
\lambda^{\overline{a}}_-
 &= \left(
           i
           \alpha_-
           h^{a \overline{c}} \,
           \overline{F}_{\overline{c}}
    \right)
    \left(
           h_{a \overline{b}} \,
           A^{\overline{b}}_{\overline{\imath} \, \overline{a}, i}
    \right)
    \psi^{\overline{\imath}}_+
    \lambda^{\overline{a}}_-
    \psi^i_+
\nonumber
\\[1ex]
 &= \psi^{\overline{\imath}}_+
    \lambda^{\overline{a}}_- \,
    A^{\overline{b}}_{\overline{\imath} \, \overline{a}, k} \,
    \left(
           i
           \alpha_-
           \psi^k_+
    \right)
    \overline{F}_{\overline{b}} \, ,           
\end{align}
where we have used
$
  F_{i \overline{\imath} a \overline{a}}
= h_{a \overline{b}} \,
  A^{\overline{b}}_{\overline{\imath} \, \overline{a}, i}  
$
in the first step.
It follows that
\eqref{delta-non-exact-2b}
cancels
\eqref{delta-non-exact-2a}
up to a total derivative, i.e.
\begin{equation}
\delta
\left(
    - \psi^i_+ 
       \lambda^a_-
       D_i F_a
\right)
  = 
  - \,
    \delta
    \left(
           \psi^{\overline{\imath}}_+
           \lambda^{\overline{a}}_- \,
           \overline{D}_{\overline{\imath}}
           \overline{F}_{\overline{a}}
    \right)
  - \alpha_-
    \partial_z \!
    \left(
           \overline{F}_{\overline{a}} \,
            \lambda^{\overline{a}}_-
    \right).  
\end{equation}
This completes our argument for the case of a holomorphic superpotential.      
\section{\label{section:Supersymmetry-nonholomorphic}Supersymmetry invariance for nonholomorphic superpotential}

In this section, we will extend the analysis of section
\ref{section:Supersymmetry-holomorphic} 
to the case in which the superpotential is not holomorphic.
This requires revisting the steps in
\eqref{delta-non-exact-2a}
and
\eqref{cancel-first-terms-rhs-nonexact-2a-2b}
where we used
$ \overline{F}_{\overline{a}, k} = 0 $.
Allowing for
$ \overline{F}_{\overline{a}, k} \ne 0 $,
\eqref{cancel-first-terms-rhs-nonexact-2a-2b}
becomes    
\begin{align*}
- \,
  \alpha_-
  \overline{F}_{\overline{a}} \,
  D_z \lambda^{\overline{a}}_-
  &= \left(
           \alpha_-
           \partial_z \phi^{\overline{\imath}}
     \right)       
     \lambda^{\overline{a}}_- \,
     \overline{D}_{\overline{\imath}} \,
     \overline{F}_{\overline{a}}
   - \alpha_-
     \partial_z \!
     \left(
            \overline{F}_{\overline{a}} \,
            \lambda^{\overline{a}}_-
     \right)     
   + \alpha_- 
     \overline{F}_{\overline{a}, k} \,
     \partial_z \phi^k
     \lambda^{\overline{a}}_- \, .    
\end{align*}
It follows that the cancellation described by
\eqref{cancel-first-terms-rhs-nonexact-2a-2b}
will still apply provided that the constraint
\begin{equation}
\label{constraint-1}
\overline{F}_{\overline{a}, k} \,
\partial_z \phi^k
\lambda^{\overline{a}}_-
  = 0     
\end{equation}
is satisfied.
Furthermore, in the next to last step of
\eqref{delta-non-exact-2a}, 
we now have
\begin{align}
\psi^{\overline{\imath}}_+
\lambda_-^{\overline{a}}
 &
\left\{
        \overline{\partial}_{\overline{\imath}}
        \left[ \, 
               \overline{F}_{\overline{a},k}
               \left( 
                      \delta \phi^k
               \right)       
        \right]       
      - A^{\overline{b}}_{\overline{\imath} \, \overline{a}}
        \left[
               \overline{F}_{\overline{b}, k}
               \left(
                      \delta \phi^k
               \right) 
        \right]  
\right\}
\nonumber
\\[1ex]
 &= \psi^{\overline{\imath}}_+
    \lambda_-^{\overline{a}}
    \left\{
           \overline{\partial}_{\overline{\imath}}
           \left[ 
                  \overline{F}_{\overline{a},k}
                  \left(
                         i
                         \alpha_-
                         \psi^k_+
                   \right)
           \right]                     
         - A^{\overline{b}}_{\overline{\imath} \, \overline{a}} \,
           \overline{F}_{\overline{b}, k}
           \left(
                  i
                  \alpha_-
                  \psi^k_+
           \right)  
    \right\}
\nonumber
\\[1ex]
 &= \psi^{\overline{\imath}}_+
    \lambda_-^{\overline{a}}
    \left(
           \overline{\partial}_{\overline{\imath}} 
           \overline{F}_{\overline{a},i}              
         - A^{\overline{b}}_{\overline{\imath} \, \overline{a}} \,
           \overline{F}_{\overline{b}, i}  
    \right)
    \left(
           i
           \alpha_-
           \psi^i_+
    \right)
\nonumber
\\[1ex]
 &= \psi^{\overline{\imath}}_+
    \lambda_-^{\overline{a}}
    \left[
           \overline{\partial}_{\overline{\imath}} 
           \overline{F}_{\overline{a}, i}              
         + A^{\overline{b}}_{\overline{\imath} \, \overline{a}, i} \,
           \overline{F}_{\overline{b}}
         - \partial_i \!
           \left(
                  A^{\overline{b}}_{\overline{\imath} \, \overline{a}} \,
                  \overline{F}_{\overline{b}}
           \right)     
    \right]
    \left(
           i
           \alpha_-
           \psi^i_+
    \right)
\nonumber
\\[1ex]
 &= \psi^{\overline{\imath}}_+
    \lambda_-^{\overline{a}}
    \left[
           \partial_i \!
           \left(
                  \overline{\partial}_{\overline{\imath}} 
                  \overline{F}_{\overline{a}}
                - A^{\overline{b}}_{\overline{\imath} \, \overline{a}} \,
                  \overline{F}_{\overline{b}}  
           \right)              
         + A^{\overline{b}}_{\overline{\imath} \, \overline{a}, i} \,
           \overline{F}_{\overline{b}}  
    \right]
    \left(
           i
           \alpha_-
           \psi^i_+
    \right)
\nonumber
\\[1ex]
 &= \psi^{\overline{\imath}}_+
    \lambda_-^{\overline{a}}
    \left(
           \partial_i
           \overline{D}_{\overline{\imath}} 
           \overline{F}_{\overline{a}}              
         + F_{i \overline{\imath} \, a \overline{a}} \,
           h^{a \overline{b}} \,
           \overline{F}_{\overline{b}}  
    \right)
    \left(
           i
           \alpha_-
           \psi^i_+
    \right),            
\end{align}
where we have used 
$ 
  A^{\overline{b}}_{\overline{\imath} \, \overline{a}, i}
= h^{a \overline{b}} \,
  F_{i \overline{\imath} \, a \overline{a}}
$  
in the last step.
It follows that, in addition to requiring
\eqref{constraint-1},
supersymmetry imposes the constraint
\begin{equation}
\label{constraint-2}
  \partial_i
  \overline{D}_{\overline{\imath}} 
  \overline{F}_{\overline{a}}              
+ F_{i \overline{\imath} \, a \overline{a}} \,
  h^{a \overline{b}} \,
  \overline{F}_{\overline{b}}
= 0 \, .             
\end{equation}
Various special cases of
\eqref{constraint-2}
were used in
\cite{GaravusoSharpe:Analogues}
to establish properties of Mathai-Quillen form analogues which arise in the corresponding heterotic Landau-Ginzburg models.
In that paper, it was claimed that supersymmetry imposes those constraints.
In this talk, we have presented details worked out in
\cite{Garavuso:Curvaure-constraints}
supporting that claim.
It would be interesting to see what constraints are imposed by supersymmetry in other models when the superpotential is nonholomorphic.
%

\end{document}